# OSIRIS-REx Operational Key Decision Points: A Retrospective

Rich Burns[a]*, Dante S. Lauretta[b]

[a] *NASA Goddard Space Flight Center, Greenbelt, MD, USA, rich.burns@nasa.gov*
[b] *Lunar and Planetary Laboratory, University of Arizona*, Tucson, AZ, USA, lauretta@arizona.edu
\* Corresponding Author

## Abstract

OSIRIS-REx, NASA's first asteroid sample return mission, rendezvoused with the near-Earth asteroid Bennu in 2018 and delivered a 121.6-gram sample to Earth in 2023 — the largest amount of material ever recovered from a planetary body beyond the Moon. The operations phase of OSIRIS-REx was considered the most challenging robotic mission that NASA had undertaken, owing to the tight constraints on spacecraft performance and the microgravity environment. In preparation for the sample collection and return campaign, the mission leadership defined four operational key decision points (OKDPs) at critical junctures: sample site selection, rehearsal and execution of the sample collection maneuver, sample stow, and Earth return. This publication examines these OKDPs in depth, inclusive of rationale, implementation, decision processes, and efficacy. We also review other decisions that enabled mission success despite the unexpectedly rugged nature of Bennu's surface. This information should be beneficial for future challenging operational concepts where science-in-the-loop, time-critical decision making is integral to success.

**Acronyms/Abbreviations**
CP - Checkpoint
MP - Matchpoint
NFT - Natural Feature Tracking
OKDP - Operational Key Decision Point
OSIRIS-REx – Origins, Spectral Interpretation, Resource Identification, and Security–Regolith Explorer
PI – Principal Investigator
PM - Project Manager
SRC - Sample Return Capsule
TAG - touch-and-go sample collection technique
TAGSAM - Touch-and-Go Sample Acquisition Mechanism
UTTR - Utah Test and Training Range

## 1. Introduction

OSIRIS-REx [1], launched in September 2016, was NASA's first asteroid sample return mission. Mission requirements called for OSIRIS-REx to rendezvous with the small, near-Earth asteroid Bennu, conduct a global mapping and characterization campaign in proximity to the small body, collect a sample from Bennu's surface, and return it safely to Earth for scientific analysis. In many respects, the mission's ambitious operations were the boldest undertaken by a NASA mission. Bennu was discovered in 1999 and characterized prior to launch [2] via Earth-based observations only. The spacecraft arrived at its target in fall of 2018 and entered orbit about Bennu on the final day of 2018. OSIRIS-REx then executed a global mapping campaign resulting in centimeter-scale characterization of Bennu's surface [e.g., 3] and millimeter-scale site-specific reconnaissance sorties for the purpose of selecting a sample collection site. Thereafter, the focus of the mission turned to sample collection via a touch-and-go (TAG) sequence, which required the spacecraft to descend to and contact the surface with its Touch-and-Go Sample Acquisition Mechanism (TAGSAM) extended on a robotic arm [4] to capture at least 60 g of pristine sample before propulsively backing away. The plan then called for OSIRIS-REx to confirm the collection of the sample by imaging the TAGSAM head and spinning the flight system with the robotic arm extended to estimate the sample mass. The sample would then be stowed in the Sample Return Capsule (SRC), the flight system would leave Bennu, and the SRC would be delivered to Earth in September of 2023.

The actual events deviated from the plan in several notable areas, which will be discussed hereafter. Regardless, the result was an unqualified success: 121.6 g of pristine Bennu sample was delivered to the Utah Test and Training Range (UTTR) on September 24, 2023. This sample mass was more than twice the mission requirement and enough to enable study of the early solar system constituents for generations to come. In fact, the early results of the sample analysis campaign have already provided key findings regarding the nature of the early solar system and potentially the delivery mechanism that seeded Earth with the building blocks of life [5,6,7].

Due to the extraordinary complexity of the operations [8] and the recognition that at certain junctures, pivotal decisions would be required that would balance competing factors related to risk, science value, and resources, the mission established a formalized decision process and four operational key decision points (OKDPs). As a Principal Investigator (PI)–led mission, the PI (Lauretta) was responsible for all decisions related to the scientific integrity of the mission as well as staying within the PI-managed cost cap and mission schedule. The PI delegated day-to-day decision authority to the Project Manager (PM; Burns). In addition, all decisions related to the safety of the flight system were the responsibility of the PM. Critical decisions related to the four OKDPs also required concurrence of NASA's Associate Administrator for the Science Mission Directorate. A Mission Planning Board, comprised of senior team members, partners, and stakeholders, was formed to analyze, deliberate, and offer recommendations [8].

While it was recognized that not all major decision points in a mission can be anticipated in advance, the four OKDPs, each representing major operational decisions, were identified prior to launch as known critical moments to the success of the mission. These are described in the sections below, followed by a discussion of other key decisions made during flight that were unknown at the time of launch.

## 2. OKDP-1: Site Selection for Reconnaissance and Sample Collection

Site selection called for the identification of the prime and backup sample collection sites based on four factors [8,9]: deliverability, safety, sampleability, and science value. Each of these factors were estimated during the global mapping campaign of Bennu to arrive at the most optimal combination of site properties. Deliverability represented the flight system's capability to arrive at the intended site based on its navigation and maneuver accuracies within the specified arrival constraints on vertical and horizontal velocity [10]. Safety projected the system's capability to contact, sample, and back away from a site safely [11]. Sampleability used information on the mean particle size and other site characteristics such as surface tilts to project the likelihood of acquiring the required 60 g of material [12]. Sampleability estimates were informed by an extensive ground test campaign using simulated regolith of varying size and composition. Science value utilized observations from the system's spectrometers to evaluate the bulk composition and variety of materials contained within the site [8,9].

Based on the Earth-based observations of Bennu, a coarse shape model was derived, and its bulk properties were inferred prior to launch [2,13]. These guided the design of the flight system, particularly the TAG guidance system and the TAGSAM [4]. The TAG guidance system was a lidar-based system intended to feedback direct measurements of the flight system's altitude on descent to the surface [14]. Due to acquisition delays of the lidar during the development of OSIRIS-REx, an optical navigation system that used the system's cameras and estimation software known as Natural Feature Tracking (NFT) was added late in the development cycle [11].

The shape model, derived from radar measurements, proved to be very accurate [2]. Figure 1 shows an example radar measurement along with a snapshot of the derived shape model, approximately 500 m in diameter with bulk features [e.g., 3]. Inferences related to the average particle size on Bennu's surface were based on measurement of Bennu's bulk thermal inertia, the rate at which its surface heats and cools. Bennu's thermal inertia is similar to areas on Earth with low average particle size, suggesting that there should be large areas of beach-like terrain on Bennu [15,16]. Therefore, the TAGSAM was designed to ingest material with a minimum dimension of no larger than 2 cm. Upon arriving at Bennu in the fall of 2018, OSIRIS-REx began observations that soon revealed the opposite; Bennu's terrain is extraordinarily rugged, globally strewn with large boulders and tilted surfaces [3,16]. Figure 2 shows one perspective of a global mosaic of Bennu's surface from the OSIRIS-REx PolyCam instrument, taken during the mission's Preliminary Survey phase [8] prior to the spacecraft entering orbit about Bennu.

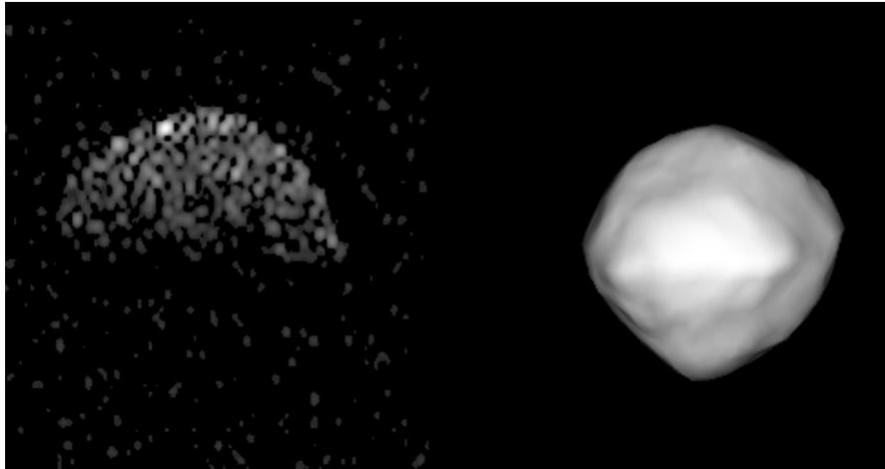
*Figure 1: Earth radar observations (left) and bulk shape model (right) derived prior to launch [2].*

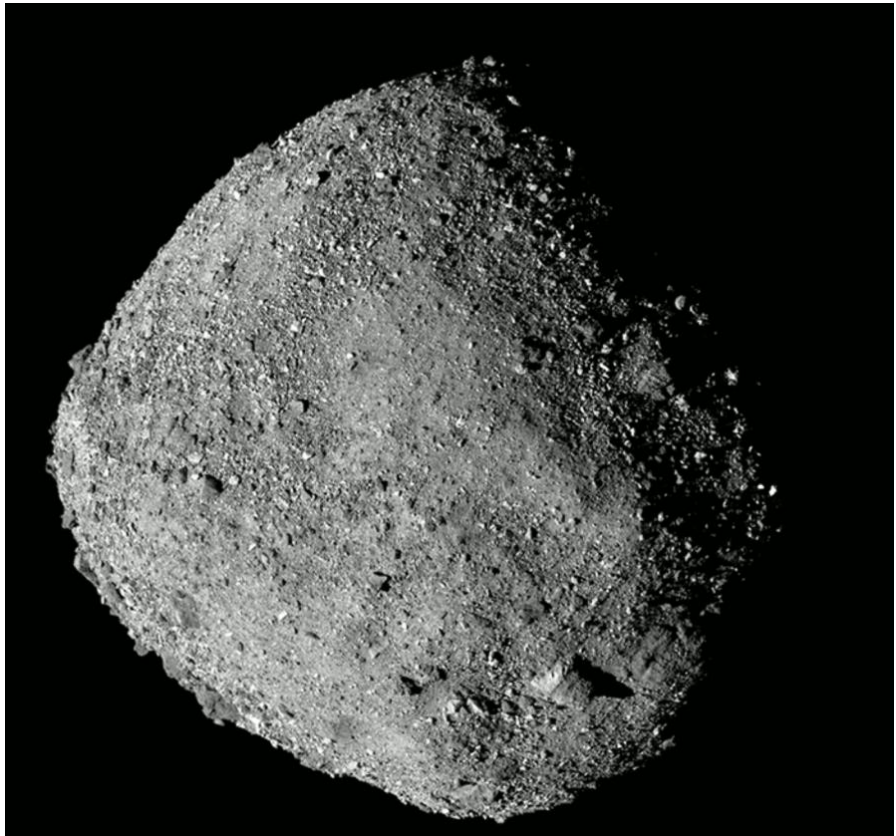
*Figure 2: PolyCam mosaic of Bennu's surface generated in late 2018.*

The unexpectedly rugged surface of Bennu presented significant challenges to the site selection process. Identification of regions that met the deliverability and safety criteria proved impossible. Most notably, the spacecraft design requirements called for it to deliver the TAGSAM head to within 25 m of a specified point on the surface [4]. This requirement was based on the inference that there would be many regions of that size on Bennu free of safety hazards (e.g., large boulders) that contained abundant fine-grained material suitable for ingestion into TAGSAM [13,2,15]. Observations from Preliminary Survey and subsequent, more detailed global measurements from the OSIRIS-REx instrument suite revealed no such area [17]. The largest hazard-free areas on Bennu are only 4 m in radius, challenging the spacecraft and navigation systems to exceed their performance requirements. Moreover, the presence of large boulders in the vicinity of candidate sites introduced a new safety concern: spacecraft tip-over causing contact with a large boulder upon the propulsive back-away maneuver, as illustrated in Figure 3. This led to the relaxation of surface tilt as a safety consideration and the adoption of hazard maps to identify regions within a site that could be hazardous in the event that spacecraft tipped as much as 25 deg from local vertical prior to backing away from the surface. The 25 deg tip-over estimate arose from refined contact dynamics analysis that modeled the spacecraft response to contact with the surface over a variety of surface characteristics.

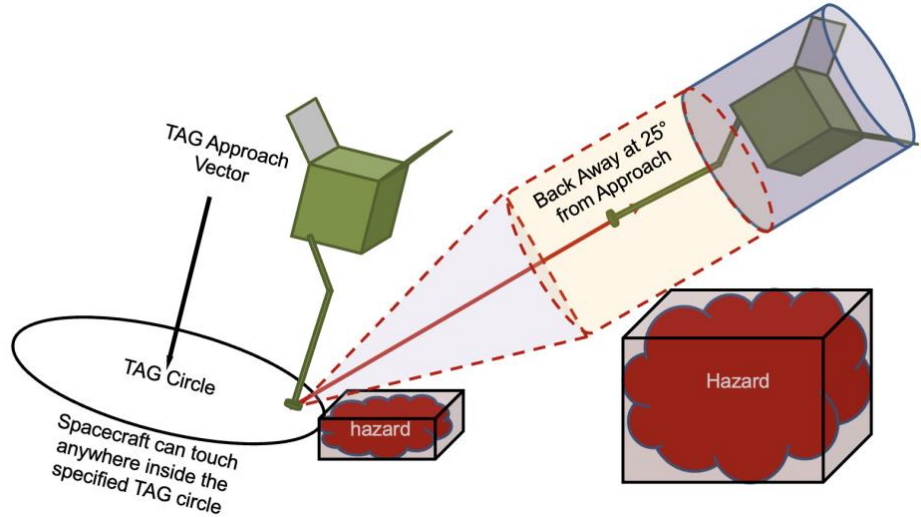

*Figure 3: Spacecraft tip-over and back-away hazards.*

Further complicating the deliverability factor was the fact that the lidar-based approach suffered from large variations in altitude measurement due to the presence of the large features that the spacecraft would fly over en route to the TAG site, leading to unpredictable results on the descent leg of the TAG trajectory. Consequently, the decision was made to adopt NFT as the on-board trajectory estimation methodology for TAG. This decision required identification of local features that could be cataloged for NFT to recognize during descent to the site [11].

By May of 2019, fifty regions of interest had been identified [8,9] through an exhaustive search of the global observations of the surface from a stable terminator orbit and from an extensive Detailed Survey campaign [8] during which the spacecraft flew on sortie trajectories to observe the surface with improved illumination conditions. After a series of site down-selections, in July 2019 the candidates were winnowed to a final four. These sites, shown in Figure 4, were now the subject of an initial reconnaissance campaign (Recon-A) to refine the characterization of each. This represented a change to the original observations to support site selection and reflected the need to further characterize sites that did not conform to the pre-launch expectations with respect to site size, safety, and sampleability. Figure 5 shows the location of each site on a projected mosaic of Bennu's surface.

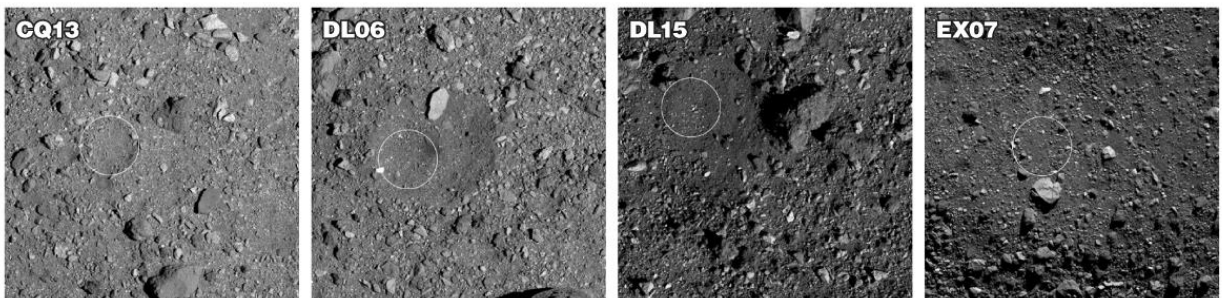

*Figure 4: The final four candidate sample sites, each shown with a representative 5-m-diameter circle centered on the site location. Initial alphanumeric site designations were ultimately replaced with avian nicknames: Kingfisher (CQ13), Osprey (DL06), Nightingale (DL15), Sandpiper (EX07).*

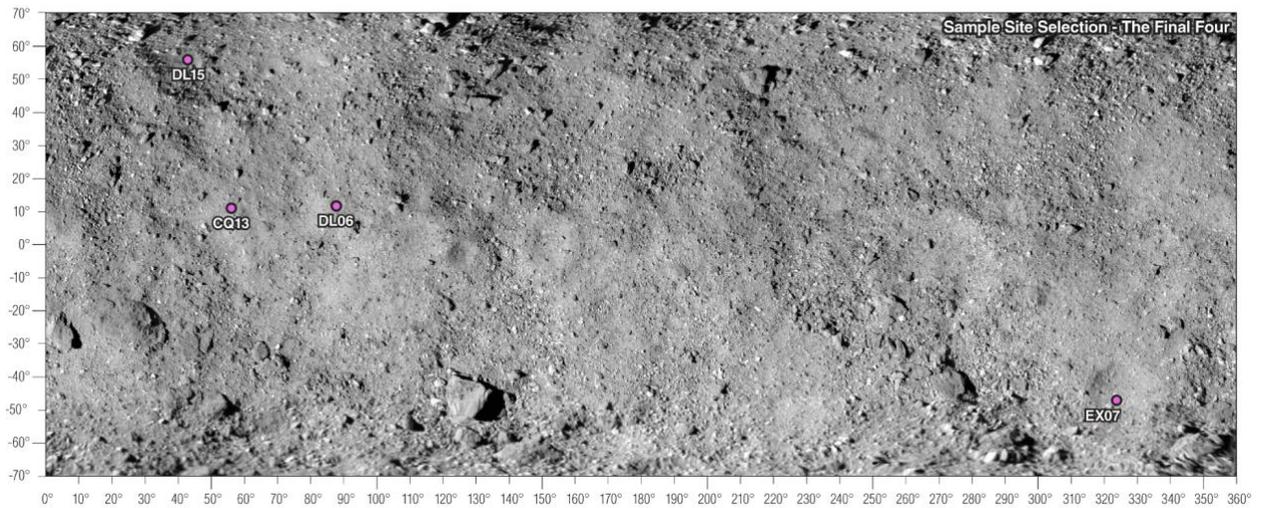

*Figure 5: The final four candidate sites, shown on the global mosaic of Bennu.*

After Recon-A observations of each of the final four candidates, the factors of site selection were now evaluated with the benefit of high-resolution, well-illuminated observations. Two sites emerged as the leading candidates. The sites were nicknamed for birds native to Egypt, consistent with the naming theme of asteroid Bennu [9]. Nightingale (internally designated DL15) and Osprey (DL06) emerged as the final two candidates on the strength of their relatively high modeled probability of collecting adequate sample and avoiding an aborted TAG attempt, respectively.

As alluded to above, the possibility of an autonomous TAG abort had been introduced as a means of mitigating the risk of spacecraft contact with boulders in proximity of the site. Note in Figure 4 that each of the final four candidate sites has these types of large features in proximity. In fact, every site considered had large boulders near or at its perimeter because the presence of large boulders constrained the allowable sample site size globally. The implementation of the mitigation took the form of a flight software patch. For each site, so-called hazard maps were derived to define regions that could be hazardous should the spacecraft touch there and tip by as much as 25 deg [9,11]. These regions could be tuned by assigning a threshold to the probability of spacecraft damage during back-away. During the descent to the site, the spacecraft updated its state using NFT (i.e., processing images, correlating the images with mapped features, and estimating position and velocity relative to them) and projected that state to the surface [11]. If the predicted point of touch was within a hazardous area at an altitude of 5 m, then the spacecraft would execute a back-away maneuver as a means of aborting the TAG attempt. Representative hazard maps are shown in Figure 6 clearly reflecting the fact that a TAG attempt at Osprey was approximately an order of magnitude less likely to result in an abort than a TAG attempt at Nightingale. For context, the estimated likelihood of a TAG abort at Nightingale based on the thresholds represented in Figure 6 was less than eight percent.

The likelihood of collecting an adequate sample was modeled as a strong function of the amount of fine-grained material that could be observed or inferred on the basis of the Recon-A observations, together with surface tilt information [9,12]. In this respect, Nightingale had a substantial advantage over Osprey, as indicated in Figure 7, which shows the stronger prevalence of high-efficiency sampleable areas, particularly near the center of the site where the point of touch was more likely to occur. Nightingale's higher-latitude location also gave it an edge in science value, as the regolith would have been exposed to less radiation, but safety, deliverability, and sampleability were necessarily the critical factors driving the prime site selection.

These competing factors argued for different outcomes. TAG at the Osprey site was less likely to result in an aborted TAG. That is, the spacecraft was more likely to touch the surface on the first TAG attempt. However, TAG at Osprey was also more likely to lead to an insufficient collected sample. While the spacecraft was designed for up to three sample attempts, several factors argued against being in the position of evaluating a potentially insufficient sample: (1) the post-sample collection imaging campaign of the TAGSAM would yield only partial information on the sample, as it was obscured by the mesh material designed to enclose it [4,18]; (2) the planned mass measurement technique had never been tested with a real sample, and its uncertainties were uncharacterized [19]; and (3) a sample attempt would likely disturb the surface features used for NFT, requiring subsequent attempts to be made at the backup. This all suggested that the likelihood of returning an insufficient sample was higher at Osprey as was the likelihood of having to repeat TAG at the backup site. However, if one assigned a low confidence level to the sampleability analysis, since its basis relied on several unconfirmable assumptions about the characteristics of the regolith [12], then there was no question that Osprey held a significant advantage with respect to likelihood of a first TAG attempt reaching the surface.

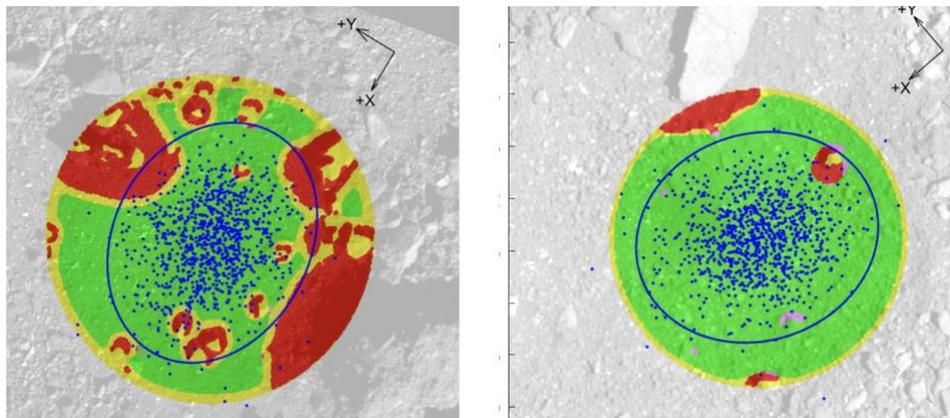

*Figure 6: Hazard maps for Nightingale (left) and Osprey (right) using a 0.5% threshold for likelihood of spacecraft damage. Red and yellow indicate areas that would result in a TAG abort. Green indicates areas free of tip-over hazards and pink indicates areas containing a hazard but one that did not meet the threshold for abort. The blue ellipses and points represent an area that contained 99% of the Monte Carlo simulations.*

As might be expected in such cases, the Mission Planning Board recommendation was split. The majority supported selection of Nightingale on the strength of higher likelihood of a sufficient sample on the first non-aborted TAG attempt, with the understanding that an aborted TAG was more likely at Nightingale than Osprey. A minority opinion emerged that identified Osprey as the preference on the basis that any TAG attempt that made it to the surface was likely to collect adequate sample regardless of the sampleability characterization. Ultimately, both the PI and the PM selected Nightingale with the concurrence of the NASA Associate Administrator for the Science Mission Directorate after a thorough briefing presented the pros and cons of each option. This was a key milestone where NASA leadership explicitly invested in the decision-making process, which was a fundamental motivation behind the definition of OKDPs: stakeholder concurrence. In the end, the preparation required to gain this level of stakeholder concurrence was likely as valuable as the concurrence itself because the decision process, data, and rationales were subjected to multiple rounds of review and iteration which ultimately produced a lasting record for future missions.

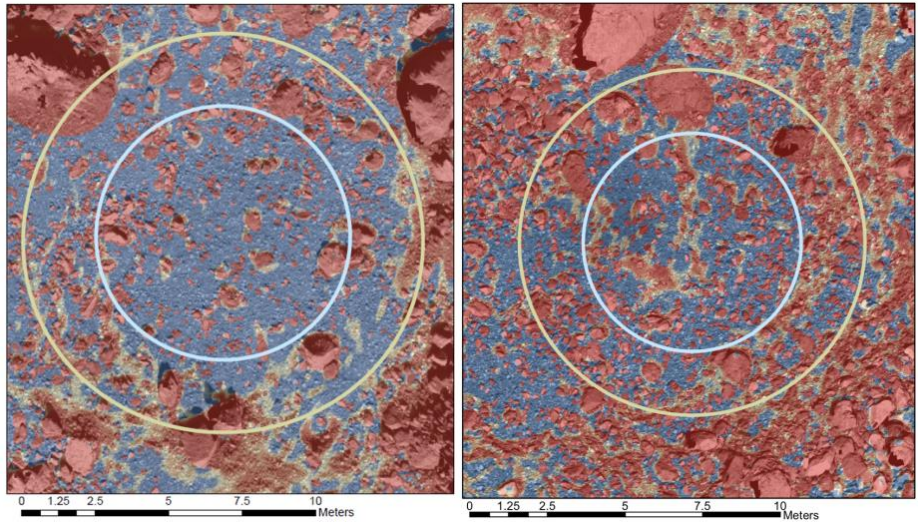

*Figure 7: Estimated sampleability efficiency for Nightingale (left) and Osprey (right) [12]. Blue indicates high efficiency while red indicates low efficiency. The tan circle denotes the area that contained 98.3% of Monte Carlo cases while the blue circle contained 80%.*

### 3. OKDP-2: Go for Rehearsal and Sample Collection

The mission design called for two rehearsals preceding TAG. These rehearsals included execution of the TAG sequence incrementally through the two maneuvers to be performed following orbit departure, namely the Checkpoint (CP) and Matchpoint (MP) maneuvers [1,8]. The TAG sequence as well as the termination of each of the rehearsals are illustrated in

Figure 8 for context.

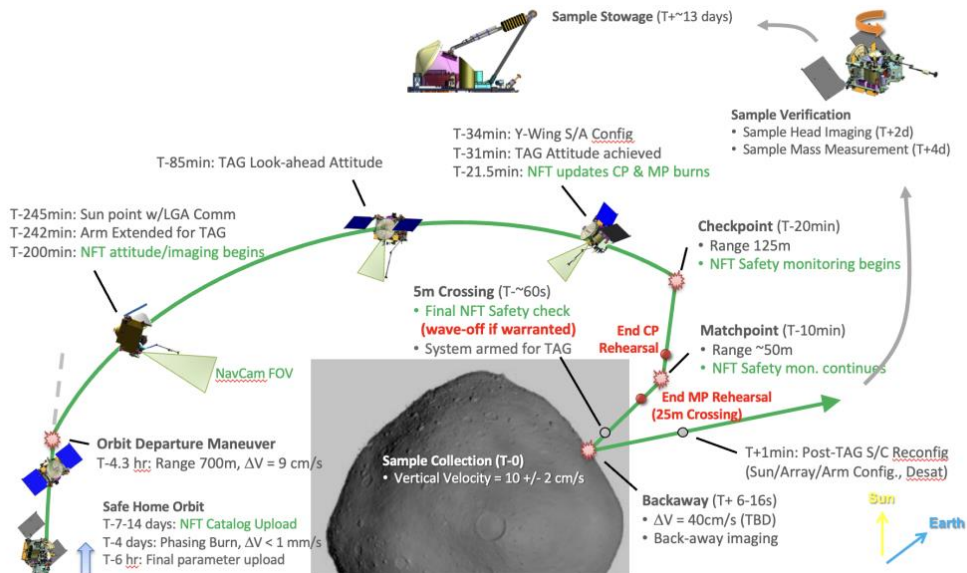

*Figure 8: TAG sequence of events and rehearsal termination points.*

Originally, OKDP-2 was envisioned prior to CP rehearsal with the idea that each rehearsal would have predefined performance criteria that could be incrementally satisfied. Should system performance remain within these bounds, a go decision for OKDP-2 would map ultimately to a go for TAG. With the change in TAG navigation approach (from lidar to NFT) and the requisite development of a NFT feature catalog, the project delayed the timing of OKDP-2 to follow CP rehearsal such that the decision could be informed by NFT performance, and the performance criteria could be more reliably bounded. Moreover, the CP rehearsal had been modified to include the CP maneuver and subsequent imagery such that evaluation of the post-maneuver NFT convergence could be assessed. Rather than OKDP-2, CP rehearsal was preceded by a comprehensive operational proficiency integrated exercise and a formal project readiness review. Additionally, NFT had been the focal point of a review by independent optical navigation experts whose recommendations had been incorporated prior to CP rehearsal. As it turned out, the performance during CP rehearsal conformed to expectations. As shown in Figure 9, CP rehearsal confirmed that NFT could correlate cataloged features and converge on updated state estimates. Crucially, it also showed that the update to the CP maneuver made a small correction toward the target, as represented by the purple ellipse, which was derived from a ground-based reconstruction and extrapolation of the remainder of the TAG sequence with realistic error sources. Notably this reconstruction benefitted from post-CP maneuver images such that a realistic post-maneuver NFT state update could be derived. The strength of the data from CP rehearsal added confidence to the OKDP-2 decision to move forward with execution of MP rehearsal and ultimately TAG. The expected performance envelope for NFT and the vehicle's maneuver capabilities had been well established.

MP rehearsal followed suit with the system entirely performing within expected bounds, setting confidence levels high for TAG, scheduled for October 20, 2020. It's notable that each rehearsal and TAG itself occurred in the heart of the Covid-19 pandemic with CP and MP rehearsals executed on April 14, 2020, and August 11, 2020, respectively. In the setting of social distancing, masking, and without a vaccine available to the general public, the team relied on the strong relationships it had developed throughout operations, but particularly in proximity operations, to navigate the pandemic's challenges.

So, as the team came together for the TAG event, confidence in the system's performance was high, yet uncertainty on Bennu's response also hung in the air. As the spacecraft descended to Bennu at a distance of nearly 2 astronomical units from Earth, the team eagerly listened for calls based on NFT updates to the likelihood of a TAG abort. As the spacecraft flew through the abort altitude (5 m) below the threshold that would have triggered a wave-off maneuver, the team awaited contact. And, at just 10 cm/s descent rate, it came nearly a minute later.

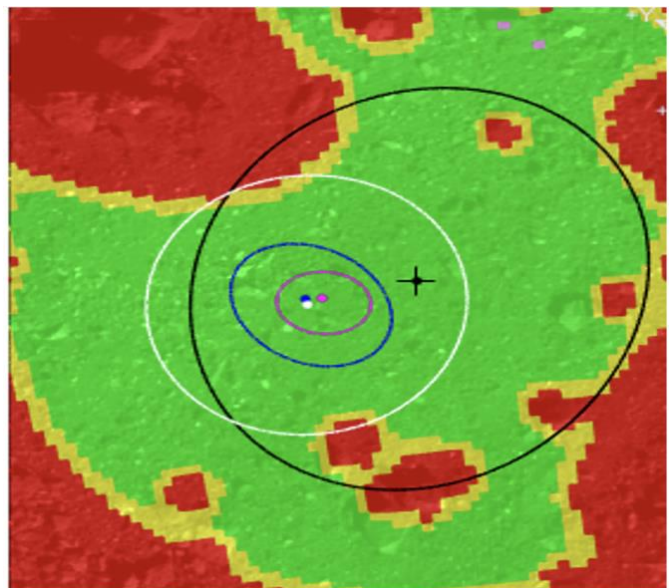

*Figure 9: CP rehearsal results: black - a priori (98.3%) error ellipse, 6.3 m SMA; blue – error ellipse based on last NFT-computed state, from ground-based analysis; white – NFT-predicted contact point and uncertainty from onboard solutions, centered 2.6 m from target; purple – expected contact point and uncertainty from ground-based reconstruct.*

TAG and successful back-away maneuvers were verified. Subsequent downlink of images as seen in Figure 10 showed a shockingly energetic impact on the surface [20]. Post-TAG reconstruction [18] estimated the TAGSAM had penetrated Bennu's surface up to nearly 0.5 m. Optimism was high that the TAGSAM had captured its sample.

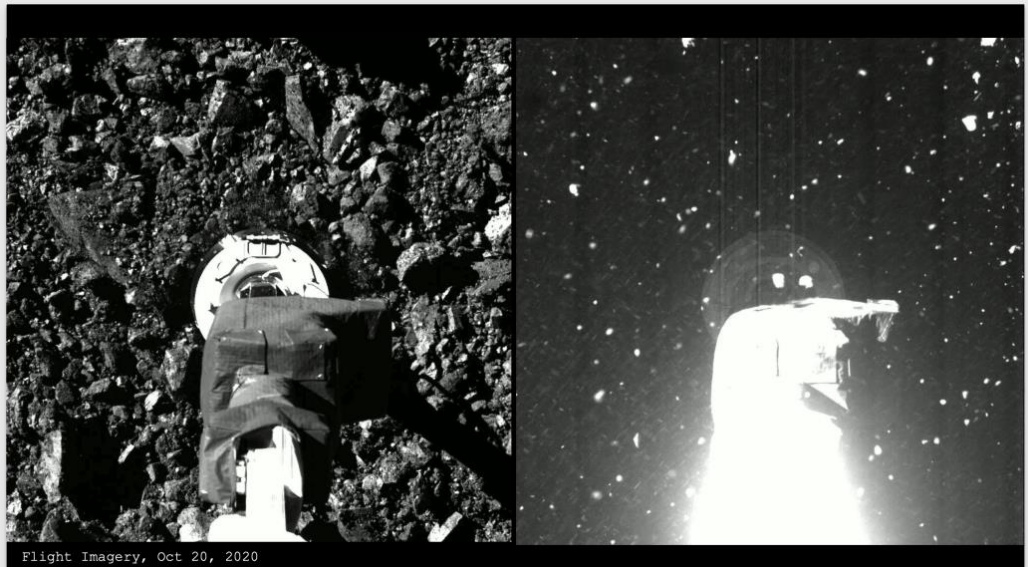

Figure 10: Images of the TAGSAM head at the moment prior to contact (left) and immediately following (right) with an enormous shower of particles liberated from the surface.

## 4. OKDP-3: Stow Sample

It's useful at this point to put OKDP-3 into the context of the process used for the first two OKDPs, which had been executed with a large amount preparation: data gathering, extensive modeling, independent reviews, exercises, etc. The overall information flow is portrayed in Figure 11. It was recognized that the decision to stow the sample would have to be made in a shorter time, nominally 13 days as shown in Figure 8, and that decision would be supported by an incomplete data set. The shorter timeline was a recognition that while the TAGSAM was designed to contain the collected sample, the sample was vulnerable while it remained unstowed. One of the threats was to sample integrity due to heating inside the TAGSAM head due to solar illumination. For this reason, TAGSAM's post-TAG exposure to the sun was limited; however, unplanned events like a safe mode entry could expose the TAGSAM inadvertently and risk heating up the sample, so time was of the essence.

TAGSAM contained no mass measurement device or internal visualization. Imagery would be taken by the spacecraft's SamCam instrument with the TAGSAM head positioned in various orientations. However, the images could only offer partial insight into the contents as they would be partially obscured by mesh material containing the sample and the transparent mylar flap that served as a check valve [18]. Analyses of the TAG imagery and inferences of the surface properties would facilitate estimates of the amount of material mobilized and its size distribution. Lastly, the spacecraft would spin with the TAGSAM arm extended as the most direct measure of sample mass. Even this measurement would have large uncertainties, though, with an estimated 3-sigma uncertainty of ±45 g compared to the 60 g sample mass requirement. Due to the relatively tight time scale to make the stow decision, "lean forward" stow criteria were presented at OKDP-2. Ultimately, the decision was made to lean toward stow if the mass measurement and other criteria represented >90% likelihood of a sample mass exceeding 60 g. Accounting for the uncertainties in the sample mass measurement, a measured value exceeding 80 g would lead to >90% confidence of a true sample mass more than 60 g [19].

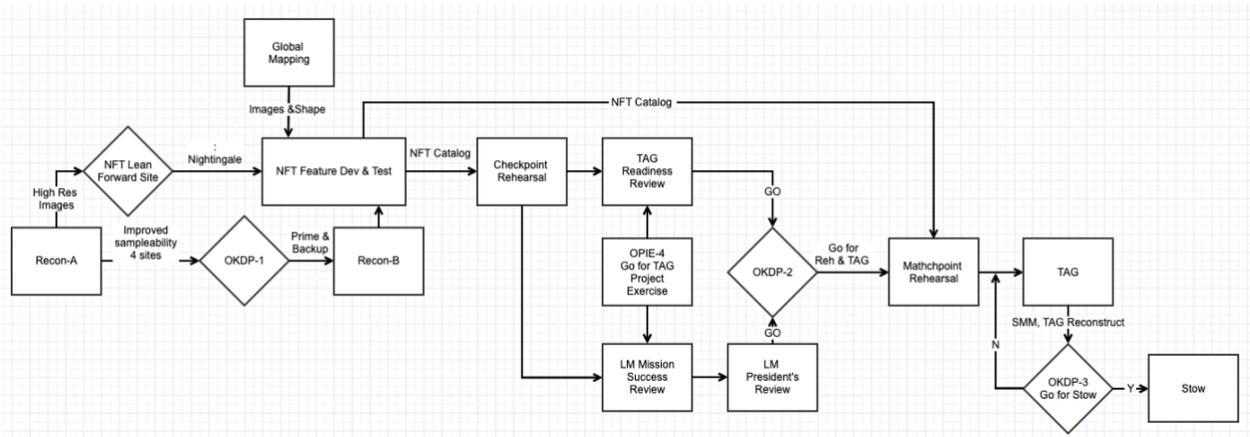
*Figure 11: OKDP information flow and process through sample stow.*

As the images from TAG began to be downlinked while the spacecraft drifted away from Bennu, it became clear that a significant amount of fine-grained material was available at the site; it also became clear that the TAGSAM head had penetrated deep (~0.5 m) into Bennu's surface, suggesting that material should have been forced into it. However, when the spacecraft began imaging the TAGSAM head two days later, two facts became evident: (1) a significant amount of sample could be seen through the mesh sides of the head and through the mylar flap, and (2) each time the head was repositioned for imaging, sample was leaking out as seen in Figure 12 due to a rock that had wedged the mylar flap open [18]. It's notable that this had never been encountered in the extensive ground test campaign [4]. Nonetheless, it was clearly a time-critical development and warranted an expeditious response.

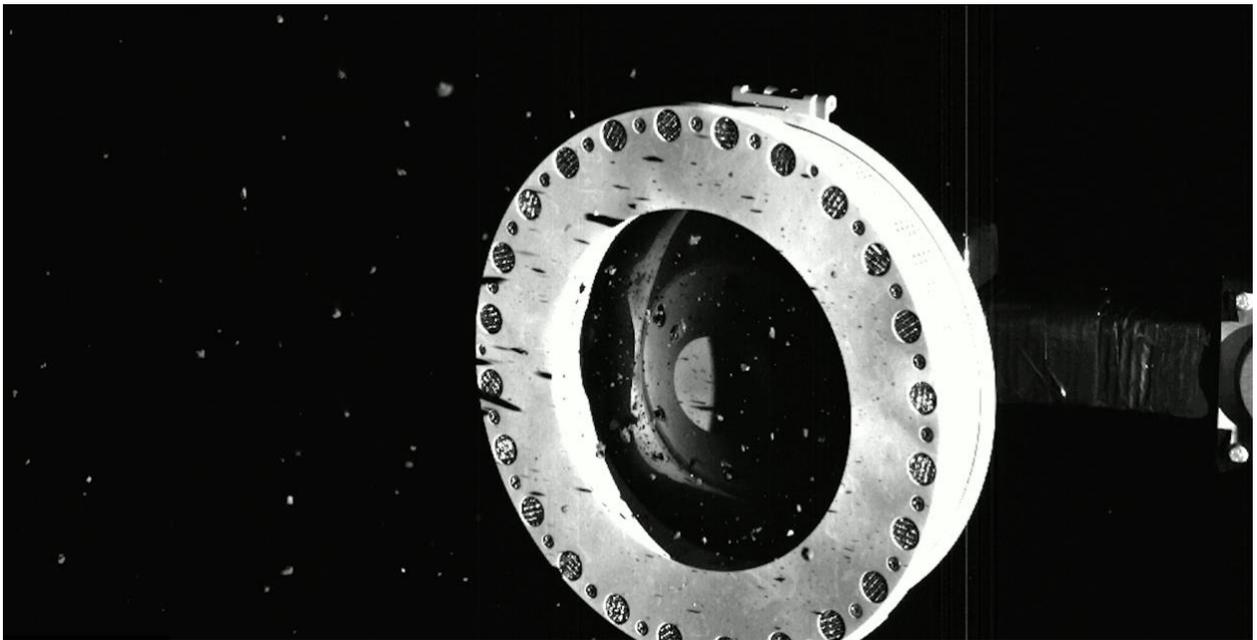
*Figure 12: Post-TAG TAGSAM imaging revealed the presence of a significant sample, though it was escaping containment in the TAGSAM head.*

The project leadership team moved swiftly toward the decision to forego the sample mass measurement activity as the motion of the TAGSAM and spacecraft involved would have certainly exacerbated the sample escape. After briefly entertaining spacecraft motions that could move the sample away from the source of the leak, leadership decided the most prudent approach was to minimize spacecraft and TAGSAM motion and move directly into the stow sequence.

With the significant data supporting the fact that a considerable amount of regolith had been collected but without definitive information on how much had escaped, the Associate Administrator concurred with the decision. The rationale included the fact that a second TAG attempt with the TAGSAM head's mylar flap wedged open would be a scenario for which we had no models or test experience to draw from in order to predict the outcome. Further, image analysis derived estimates of the escaped sample mass that could be seen in the images. The mass of the observable, escaped material was judged to be small even when multiplied by the number of TAGSAM motions used in the imaging campaign and those used to stow the sample. In the end, the stow sequence executed nominally and the TAGSAM head was safely enclosed in the SRC awaiting its ride back to Earth.

5. **OKDP-4: Earth Return**

The transit back to Earth began with an asteroid departure maneuver on May 10, 2021. On September 24, 2023, the SRC, shown with the flight system in Figure 13, was released from the spacecraft just four hours in advance of Earth atmospheric entry [5]. After release, the spacecraft performed a divert maneuver to avoid Earth, and the SRC continued on a ballistic trajectory toward UTTR [20]. The notable attribute about Earth return was that it was the only operation of the mission that had to happen at an exact time. The SRC release had to happen at its designated time in order to target UTTR; failure to release at that time meant the spacecraft would have retained the SRC and entered a heliocentric orbit with perihelion solar range well below solar ranges that OSIRIS-REx was designed to operate within, both delaying and seriously jeopardizing mission success. Moreover, the SRC release decision criteria related to flight system readiness and ground personnel safety rather than science value, which also distinguished it from the prior OKDPs. As such, the decision criteria could be pre-defined such that the only remaining variables were updates to the trajectory and predicted landing area, simplifying the decision process which ultimately executed at approximately 2 AM local time. So long as the flight system was capable of releasing the SRC and the predicted SRC landing location remained within a pre-defined area at UTTR, the decision was to release. The latter criteria represented a simplified, yet more challenging criteria than NASA's previous sample return mission at UTTR, which was Stardust in 2004. For the Stardust mission, the allowable landing area encompassed the entirety of the range's restricted air space, shown as the blue polygon in Figure 14, exclusive of populated and sensitive areas. OSIRIS-REx had planned for the same allowable landing area; however, the range regulations had changed in the interim to be more constraining.

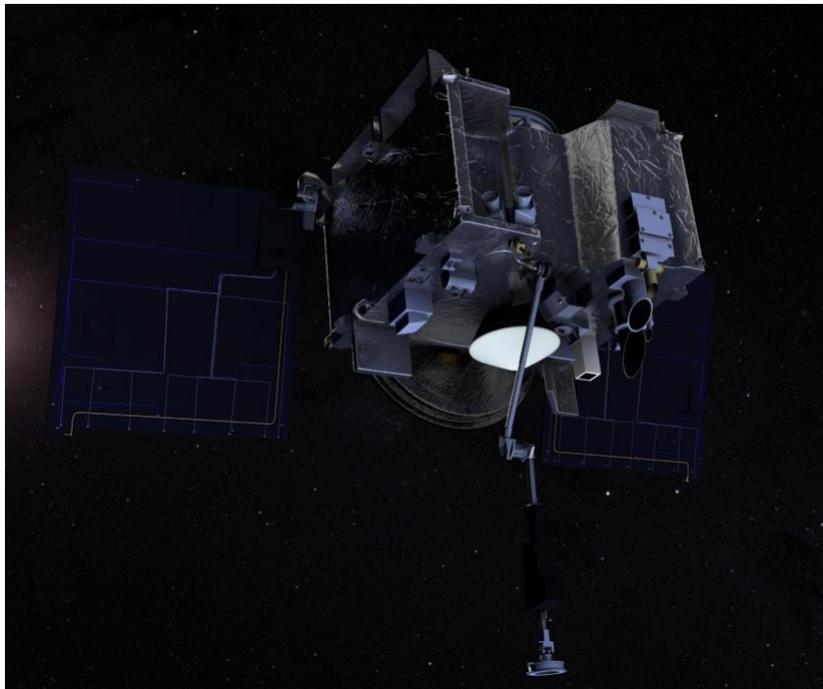

*Figure 13: Illustration of the OSIRIS-REx flight system with the SRC (heat shield in white) and the TAGSAM head at the end of the extended robotic arm.*

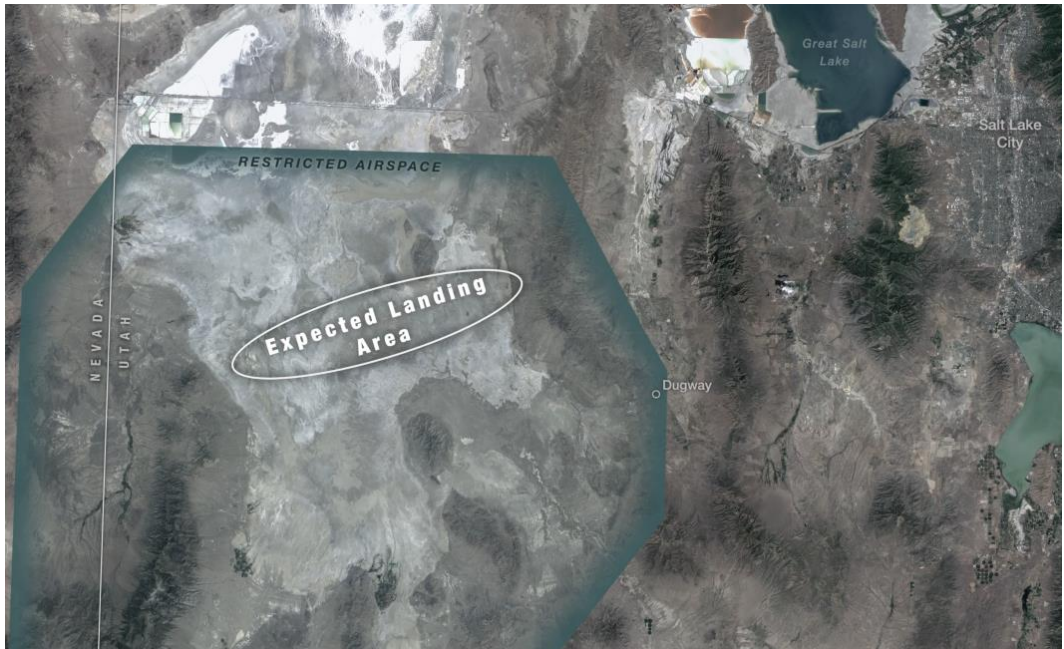
*Figure 14: Location of the SRC expected landing area relative to Salt Lake City.*

The reference landing ellipse, which was used to verify the SRC design requirements and represented in the original Inter-Agency Agreement with UTTR, formed the expectation of UTTR representatives about what areas they would need to close and secure. Ultimately, a slightly larger ellipse, labeled in Figure 14 as the expected landing area, accommodated the few, off-nominal reentry scenarios that were considered credible by the project. While the more constrained allowable footprint required a significant and unexpected verification effort to show that the SRC release conditions mapped into safe landing areas under a variety of atmospheric conditions, it ultimately simplified the OKDP-4 decision process significantly. The robust verification effort led to high confidence in the mapping between SRC release initial conditions and landing uncertainty and further that those uncertainties were contained within the allowable landing area in all but the most degenerate cases (e.g., missed maneuvers preceding SRC release) that could be ruled out in the days before the decision. Of course, ultimately the spacecraft trajectory and state of health were both nominal, enabling an easy, if early-morning decision leading to the successful return of NASA's first asteroid sample [21].

## 6.  An unexpected key operational decision

In 2019, shortly after the spacecraft entered orbit about Bennu, the team discovered evidence in imagery that Bennu was an active asteroid. That is, particles were being ejected from the surface [e.g., 22]. While the mission had scanned the vicinity of Bennu for natural satellites during its approach and found none [23], no one had expected that Bennu would be actively shedding material [13]. Yet images taken from Bennu orbit were now showing clear evidence that particles lofted from the surface were both escaping Bennu orbit and being redeposited back on Bennu's surface. Figure 15 shows a composite image of Bennu and particles being ejected from its surface. Of course, the immediate concern was whether the flight system was safe in orbit with an unknown flux of particles emanating from the surface. The penalty for leaving orbit was severe in terms of retaining a schedule to support the site selection mapping campaign and ultimately TAG within the resources of the project. The team quickly pivoted to a safety analysis which used all available images to constrain the observable characteristics of the particles and the frequency at which ejection events occurred on Bennu. These constraints allowed for the establishment of upper limits on the observed particle energy and flux that facilitated reasonable safety risk estimates, which showed that the likelihood of a particle strike was small and that the consequence of the low-energy impact would also likely be insignificant. This quick turn-around assessment and decision to remain in orbit allowed the mission to stay on track toward sample collection at an acceptable risk posture.

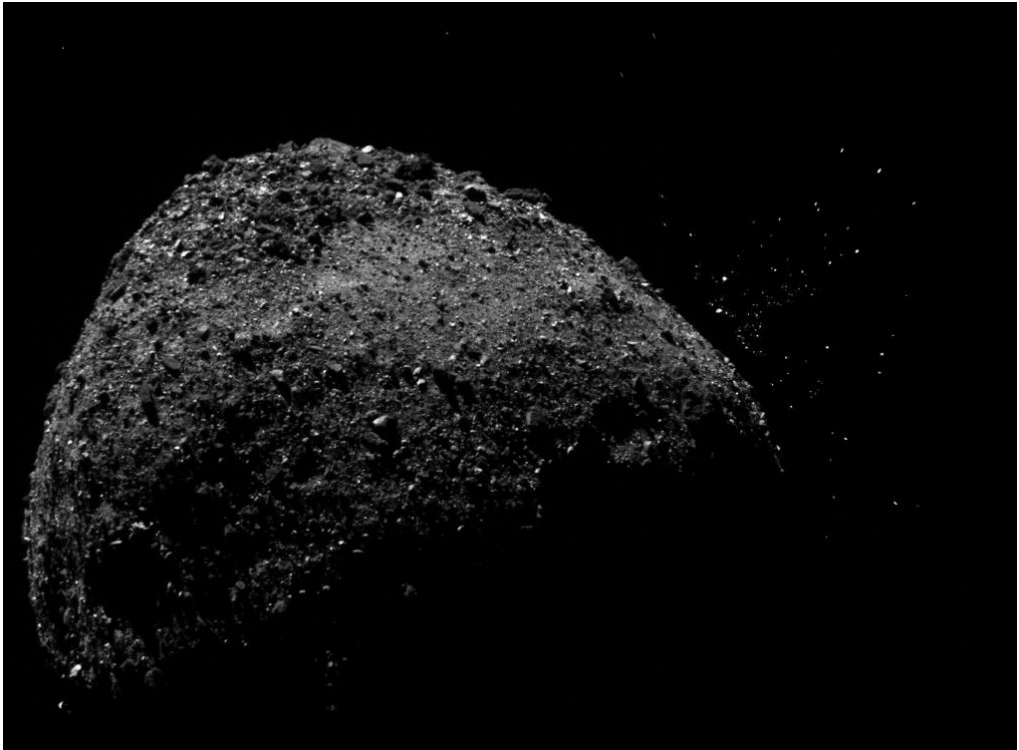
*Figure 15: Composite image of Bennu and particles ejecting from its surface [22].*

**6. Conclusions**

Operational decisions are made every day for all space missions, yet some decisions can have irreversible and profound effects both positive and negative. These decisions warrant forethought, not just on the outcome, but the process, data, and criteria to be applied. For OSIRIS-REx, the definition of these key turning points in the exploration of the small body Bennu, and the preparation and discussion of decision criteria, led to clear communications, improved stakeholder investment, and ultimately a highly successful outcome. Not always did the decisions manifest in the expected manner, but the value of the plan is in the planning. Such is the nature of exploration.

**Acknowledgements**

This material is based upon work supported by NASA under Award NNH09ZDA007O and Contract NNM10AA11C issued through the New Frontiers Program. The decisions articulated here are all attributable to the dedicated efforts of the amazing OSIRIS-REx team, whose total commitment to the mission made the difference every single day, regardless of whether that day contained an OKDP.